\def\etal{{\it et al.}}

\def\fm3{fm$^3$}
\def\fmm3{fm$^{-3}$}

\def \ba88{{\it Particles and Fields 3} (Proceedings of the 1988 Banff Summer
Institute on Particles and Fields), edited by A. N. Kamal and F. C. Khanna
(World Scientific, Singapore, 1989)}

\def \be87{{\it Proceedings of the Workshop on High Sensitivity Beauty
Physics at Fermilab,} Fermilab, Nov. 11-14, 1987, edited by A. J. Slaughter,
N. Lockyer, and M. Schmidt (Fermilab, Batavia, IL, 1988)}

\def \cp89{{\it CP Violation,} edited by C. Jarlskog (World Scientific,
Singapore, 1989)}
\def \dpf91{{\it The Vancouver Meeting - Particles and Fields '91}
(Division of Particles and Fields Meeting, American Physical Society,
Vancouver, Canada, Aug.~18-22, 1991), ed. by D. Axen, D. Bryman, and M. Comyn
(World Scientific, Singapore, 1992)}

\def \hb87{{\it Proceeding of the 1987 International Symposium on Lepton and
Photon Interactions at High Energies,} Hamburg, 1987, ed. by W. Bartel
and R. R\"uckl (Nucl.~Phys.~B, Proc. Suppl., vol. 3) (North-Holland,
Amsterdam, 1988)}

\def \ite{{\it et al.}}

\def \ky85{{\it Proceedings of the International Symposium on Lepton and
Photon Interactions at High Energy,} Kyoto, Aug.~19-24, 1985, edited by M.
Konuma and K. Takahashi (Kyoto Univ., Kyoto, 1985)}
\def \lat90{{\it Results and Perspectives in Particle Physics} (Proceedings of
Les Rencontres de Physique de la Vallee d'Aoste [4th], La Thuile, Italy, Mar.
18-24, 1990), edited by M. Greco (Editions Fronti\`eres, Gif-Sur-Yvette,
France,
1991)}
\def \lg91{International Symposium on Lepton and Photon Interactions, Geneva,
Switzerland, July, 1991}
\def \lkl87{{\it Selected Topics in Electroweak Interactions} (Proceedings of
the Second Lake Louise Institute on New Frontiers in Particle Physics, 15 --
21 February, 1987), edited by J. M. Cameron \ite~(World Scientific, Singapore,
1987)}

\def \np#1#2#3{{\it Nucl. Phys.} {\bf#1} (#3) #2}
\def \oxf65{{\it Proceedings of the Oxford International Conference on
Elementary Particles} 19/25 Sept.~1965, ed.~by T. R. Walsh (Chilton, Rutherford
High Energy Laboratory, 1966)}

\def \pl#1#2#3{{\it Phys. Lett.} {\bf#1} (#3) #2}

\def \pr#1#2#3{{\it Phys. Rev.} {\bf#1} (#3) #2}

\def \si90{25th International Conference on High Energy Physics, Singapore,
Aug. 2-8, 1990, Proceedings edited by K. K. Phua and Y. Yamaguchi (World
Scientific, Teaneck, N. J., 1991)}
\def \slac75{{\it Proceedings of the 1975 International Symposium on
Lepton and Photon Interactions at High Energies,} Stanford University, Aug.
21-27, 1975, edited by W. T. Kirk (SLAC, Stanford, CA, 1975)}
\def \slc87{{\it Proceedings of the Salt Lake City Meeting} (Division of
Particles and Fields, American Physical Society, Salt Lake City, Utah, 1987),
ed. by C. DeTar and J. S. Ball (World Scientific, Singapore, 1987)}
\def \smass82{{\it Proceedings of the 1982 DPF Summer Study on Elementary
Particle Physics and Future Facilities}, Snowmass, Colorado, edited by R.
Donaldson, R. Gustafson, and F. Paige (World Scientific, Singapore, 1982)}
\def \smass90{{\it Research Directions for the Decade} (Proceedings of the
1990 DPF Snowmass Workshop), edited by E. L. Berger (World Scientific,
Singapore, 1991)}

\def \tasi90{{\it Testing the Standard Model} (Proceedings of the 1990
Theoretical Advanced Study Institute in Elementary Particle Physics),
edited by M. Cveti\v{c} and P. Langacker (World Scientific, Singapore, 1991)}

\documentstyle[12pt]{article}
\begin{document}
\begin{titlepage}
\begin{center}
Tel Aviv University Preprint TAUP 2256-95, \\
Weizmann Institute Preprint WIS-95/26/Jun-PH,\\
Submitted to Zeitschrift fur Physik A, Hadrons and Nuclei\\
Archive hep-ph@xxx.lanl.gov, HEPPH-9510xxx\\
Oct. 1995\\
\vspace{0.2in}
{\Large\bf How to Search for Pentaquarks\\ in High Energy Hadronic
Interactions\\}
\vspace{0.2in}
{\bf      M. A. Moinester, D. Ashery \\}
R. \& B. Sackler Faculty of Exact Sciences,
School of Physics\\
Tel Aviv University, 69978 Ramat Aviv, Israel\\
{\bf L. G. Landsberg,\\}
Institute for High Energy Physics, 142284 Protvino, Russia\\
{\bf H. J. Lipkin,\\}
Department of Particle Physics, Weizmann Institute of Science,
76100 Rehovot, Israel\\
\vspace{0.2in}
{\Large\bf  Abstract}
\end{center}
\normalsize
 The strange-anticharmed Pentaquark is a $uud\bar{c}s$ or $udd\bar{c}s$
five-quark baryon that is expected to be either a narrow resonance, or possibly
even stable against strong and electromagnetic decay. We describe this hyperon
here, its structure, binding energy and lifetime, resonance width, production
mechanisms, production cross sections, and decay modes. We describe techniques
to reduce backgrounds in search experiments and to optimize the conditions for
Pentaquark observation. Possibilities for enhancing the signal over background
in Pentaquark searches are investigated by examining predictions for detailed
momentum and angular distributions in multiparticle final states. General
model-independent predictions are presented as well as those from two models: a
loosely bound $D_{s}^-N$ molecule and a strongly-bound five-quark system.
Fermilab E791 data, currently being analyzed, may have marginal statistics for
showing definitive signals. Future experiments in the spirit of the recent
CHARM2000 workshop, such as FNAL E781 and CERN CHEOPS with $10^6-10^7$
reconstructed charmed baryon events, should have sensitivity to determine
whether or not the Pentaquark exists.
\end{titlepage}
\section*{1.~~Introduction}
Ordinary hadrons are mesons or baryons, whose quantum numbers can be described
by quark-antiquark or three-quark configurations. Unusual hadrons that do not
fit this picture would constitute new forms of hadronic matter - exotic
hadrons. Such hadrons may have valence multiquark configurations such as
$qq\bar{q}\bar{q}$ and $qqqq\bar{q}$. Exotic hadrons can have anomalous quantum
numbers not accessible to three-quark or quark-antiquark structures (open
exotic states) or even usual quantum numbers (cryptoexotic states).
Cryptoexotic hadrons can be identified only by their unusual dynamical
properties (anomalously narrow decay widths, anomalous branching ratios, etc.).
The discovery of exotic hadrons would have far-reaching consequences for
quantum chromodynamics, for the concept of confinement, and for specific models
of hadron structure (lattice, string and bag models). Detailed discussions of
exotic hadron physics can be found in recent reviews \cite
{lgl,pet,ams,hert,dov}.

We consider here possible exotic hadronic states which contain quarks with four
different flavors. The states have quark configurations $Qq\bar{q}\bar{q}$ and
$qqqq\bar{Q}$, with one heavy quark Q(c,b) and also lighter quarks q(u,d,s).
Their properties follow from the general hypothesis of ``flavor antisymmetry"
\cite {fa}, by which quark systems characterized by the maximum possible
antisymmetry of quark flavors (both quarks and antiquarks) are the most
strongly bound. For instance,this means that the $u\bar{u}d\bar{s}$ system with
ordinary $d\bar{s}$ flavors would be more bound than the $uu\bar{d}\bar{s}$ one
 with exotic values of charge $Q_c=+2$,   etc. Jaffe \cite {jaf} predicted in
this spirit that for dibaryons with six light quarks, the most bound is the
Hexaquark H = [uuddss] combination, for which not more than two quarks are in
states with identical flavors.

Thus, for the mesons and baryons with three types of light quark constituents
(u,d,s), the ``flavor antisymmetry" principle predicts that the states with
open exotic charges and flavors are not so strongly bound and may have decay
widths too large to be really observable. This  may explain why the main
candidates for light quark exotics have cryptoexotic characteristics or are
states with exotic $J^{PC}$ values, rather than having open exotic flavors such
as mesons with $Q_c= \pm 2$, or $S= \pm 2$, baryons with $S>0$, or $Q_c>2$ and
so on.

The situation can change for the exotic hadrons with four different quark
constituents $u,d,s,c$ or $u,d,s,b$, etc. For these states the flavor
antisymmetry principle allows the existence of strongly bound states with open
exotic charges and flavors. Thus,  Lipkin \cite {lip} and Gignoux \etal ~ \cite
{gig} showed that 5-quark ``anticharmed" baryons (Pentaquarks) of the P$^0$ =
[$uud\bar{c}s$] and P$^-$ = [$udd\bar{c}s$] type, or analogous "anti-beauty"
baryons, are the most bound in the 5-quark sector. There are also predictions
\cite {fa, zr} for the most bound tetraquark exotic meson, the
$\tilde{F}_s$=[$cs\bar{u}\bar{d}$]. The present report focusses on such exotic
states, with one heavy quark. At the CHARM2000 workshop \cite {kk}, a shorter
version of the present work was presented \cite {moi}. The properties of exotic
pentaquark baryons [$qqqq\bar{q},sqqq\bar{q},ssqq\bar{q},sssq\bar{q},
ssss\bar{q},qqqs\bar{s},qqss\bar{s}$] have been also discussed by Kaidalov,
Grigoryan, Ferrer, Strakovsky, and others \cite {kg,fer,ks} (see also
\cite{lgl} and references therein).

\section*{2.~~Pentaquark Binding Energy}
A very interesting situation can be realized, if there are exotic hadrons with
heavy quarks, which are the bound states of known (quasi)stable particles, with
masses which are below the threshold for strong decay. For example, if the
Pentaquark is a bound state $(ND^-_s)$ with the mass $M(P)~<~M(D^-_s)~+~M(N)$,
such a state would decay only via weak interactions and would be quasistable.
On the other hand, a resonance $(ND^-_s)$-state with a mass above the strong
threshold, would be a short-lived state that decays strongly as
$P~\rightarrow~N~+~D^-_s$.

The binding potential of a system is given by the difference between the
Color Hyperfine (CH) interaction in the system and in the lightest
color-singlet combination of quarks into which it can be decomposed. The
wave function of the H may be written as:
\begin{equation}
 \Psi _{H} = \alpha_{1}  \Psi _{6q} + \beta_{1}
 \Psi _{(\Lambda\Lambda)} + \gamma_{1}  \Psi _{(\Sigma^-\Sigma^+)}
+ \delta_{1}  \Psi _{(\Xi^-p)}.
\end{equation}
The lightest color singlet combination is the $\Lambda\Lambda$ system at 2231
MeV. The CH contribution to the binding energy of the H is about 150~MeV \cite
{jaf} in simple models of the CH interaction. Similarly, the $P^{0}$ and P$^-$
wave functions can be written as:
\begin{equation}
 \Psi _{P^0} = \alpha_{2}  \Psi _{5q} + \beta_{2}
 \Psi _{(D_{s}^-p)} + \gamma_{2}  \Psi _{(\Sigma^{+}D^{-})}
+ \delta_{2}  \Psi _{(\Lambda\bar{D}^{0})},
\end{equation}
\begin{equation}
 \Psi _{P^-} = \alpha_{3}  \Psi _{5q} + \beta_{3}
 \Psi _{(D_{s}^-n)} + \gamma_{3}  \Psi _{(\Sigma^{-}\bar{D^{0}})}
+ \delta_{3}  \Psi _{(\Lambda D^{-})}.
\end{equation}
Here, the lightest color singlet is the $D_{s}^-N$ system at 2907 MeV. The CH
contribution to the mass splitting M($D_{s}^-p$) - M(P$^0$) is the same as for
the H particle, again in simple models of the color hyperfine interaction \cite
{lip,gig}. The anti-Pentaquark is defined similarly. In general, whatever can
be said about the Pentaquark holds true also for the charge-conjugate
particles.

The calculations of Ref. \cite {fle} account for the $SU(3)_{F}$ breaking.
It was shown that as the symmetry breaking increases, the P always retains
a larger binding potential than the H, and that the binding can be several
tens of MeV. The total binding energy includes the internal kinetic energy.
Because the c quark is massive, the kinetic energy in the P is smaller than
in the H by about 15 MeV. This improves the prospects of the P to be bound.

More recently, Takeuchi, Nussinov and Kubodera \cite {tnk} studied the
effects on the Pentaquark and Hexaquark systems of instanton induced
repulsive interactions for three quarks in flavor antisymmetric states.
They claim in this framework that both Pentaquark and Hexaquark are not
likely to be bound. Also, Zouzou and Richard \cite {zr} reconsidered
previous bag model calculations for the tetraquark and pentaquark. Their
new calculation has weaker chromomagnetic attractions at short distances
and a larger bag radius for multiquark states compared to ordinary hadrons.
They find that the Pentaquark is unbound by 80 MeV, while the $\tilde{F}$
tetraquark is unbound by 230 MeV. Similar conclusions for the P and H were
given by Fleck \etal ~ \cite{fle}. Riska and Scoccola \cite {rs} recently
described the Pentaquark in a soliton model, using different parameter sets.
One set gives a bound state, while another gives a near threshold resonance.
Chow \cite {cho} discusses the $qqq\bar{c}s$ Pentaquark in the framework of the
binding of a heavy meson to a chiral soliton. Riska and Scoccola  \cite {rs}
and Oh et al. \cite {oh}  discuss the properties of a heavy $qqqq\bar{c}$
Pentaquark without strangeness. Shmatikov \cite {shma} discusses bound
pentaquarks with a molecular type baryon-meson structure, including ND$^*$ and
NB$^*$.

A very weakly bound $D_{s}^-$p deuteron-size bound state just below threshold
with a structure very different from that of the strongly bound proton size
Pentaquark might still be consistent with these recent calculations,
considering all the model uncertainties. The $D_{s}^-$p system does not have
Pauli blocking and repulsive quark exchange interactions which arise in all
hadron-hadron systems where quarks of the same flavor appear in both hadrons.
Thus, even a comparatively weak short range interaction could produce a
relatively large size bound state analogous to the deuteron, with a long
$D_{s}^-$p tail in its wave function and a good coupling to the $D_{s}^-$p
system. Because in the Pentaquark, unlike the deuteron, there is no short range
repulsion, its structure at short distances will be quite different from that
of the deuteron. This component too has it's influence on the production
mechanism, as discussed in subsection 4.2. The deuteron-like state will be
stable against strong and electromagnetic decays. Since the $D_{s}^-$p pair has
some 50-75 MeV lower mass than other meson-baryon cluster components in the
Pentaquark, it will be the dominant component in a weakly bound deuteron-like
Pentaquark. Considering all the uncertainties in knowing the Pentaquark binding
energy, our experimental approach is to search for both strongly and weakly
bound Pentaquarks, as well as unbound Pentaquark resonances.

\section*{3.~~Pentaquark Structure and Decay Modes}
   There are different possibilities for the internal structure of observable
(not very broad) exotic hadrons. They can be bound states or near threshold
resonance structures of known color singlet sub-systems (e.g. $\Lambda \Lambda$
for the H \cite {mm} or $D_{s}^{-}p$ for the P$^0$). But they can have more
complicated internal color structure, such as baryons with color octet and
sextet bonds [$(qqq)_{\bar{8c}} \times (q\bar{q})_{8c}$] and
[$(qq\bar{q})_{\bar{6}c} \times (qq)_{6c}$] (see ref. \cite{A}). We designate
all such structures as direct five quark configurations. If color substructures
are separated in space by centrifugal barriers, then exotic hadron resonances
can have not very broad or even anomalously narrow decay widths, because of
complicated quark rearrangements in the decay processes. If these exotic
hadrons are bound strongly, they can be quasistable, with only weak decays.

The wave function of the Pentaquark may contain two-particle cluster
components, each corresponding to a pair of known color singlet particles; and
also a direct five quark  component. The Pentaquark production mechanism and
its decay modes depend on these components. The $P^{\circ}$ can be formed for
example by the coalescence of $pD^{-}_{s}, \Lambda \overline{D}^{\circ}, p
D_{s}^{*-}, \Sigma^{+} D^{-} + \Sigma^{\circ} \overline{D}^{\circ}, \Lambda
\overline{D}^{*\circ}, \Sigma^+ D^{*-} + \Sigma^{\circ} \overline{D}^{*\circ}$;
or by a one-step hadronization process. Let us consider three color-singlet
components of the $P^{0}: D^-_{s}p$ (2907 MeV), $D^-\Sigma^+$ (3058 MeV) and
$\overline{D^{0}}\Lambda$ (2981 MeV). The relative strengths of these
components depend strongly on the binding energy, as discussed above for the
deuteron-like Pentaquark. Pentaquark searches in progress in E791 \cite
{791,jl791} are based on the charged particle decay components of different
Pentaquark decay modes: $D_{s}^-p \rightarrow \phi \pi^- p$ (B=3.5\%),
$D_{s}^-p \rightarrow K^{*0}K^-p$ (B=3.3\%), $D^{-} \Lambda \rightarrow
K^+\pi^-\pi^-\Lambda$ (B=8\%), $\overline{D^{0}}\Lambda \rightarrow
K^-\pi^+\Lambda$ (B=4\%) and $\overline{D^{0}}\Lambda \rightarrow
K^-\pi^+\pi^+\pi^-\Lambda$ (B=8\%). The indicated branching ratios are those of
the on-shell D-meson. Such Pentaquark branching ratios are plausible in a model
where the D meson decays weakly, while the proton and $\Lambda$ act as
spectators. Weak decays of virtual color singlet substructures in bound states
are possible, $\Lambda D^0$ or $\Sigma^+ D^-$ for example, if their masses are
smaller than the $D_{s}^-$p threshold. In other cases,  there would be strong
decays through quark rearrangement ($\Sigma^+ D^-)_{bound} \rightarrow D_{s}^-
+ p$, and so on. Even if the masses are smaller, the phase space favors decay
to the lightest system. The phase space factor would cause the partial width
for any decay mode to be smaller than for the on-shell decay, making the total
lifetime longer.

The decay through the direct five-quark component may open many additional
channels, which may shorten the lifetime of the Pentaquark and reduce the
experimental possibilities to observe it. The direct five-quark P$^0$ component
may for example decay weakly via Cabibbo allowed  or suppressed direct  or
exchange diagrams. As a result, the P$^0$ may decay into $\pi^- p$, $K^- p$,
$\Sigma^- K^+$, $\Sigma^+ K^-$, $\Sigma^+ \pi^-$, $\Sigma^- \pi^+$, etc. Each
of these decays may have one or more $\pi^+ \pi^-$ pairs, in addition to the
particles shown. The observation of such decay channels would give important
information on the Pentaquark internal structure. These decays may be observed,
if the branching ratios are not too small. Depending on the decay mechanisms,
the Pentaquark lifetime may then be shorter or longer than the $D_s$ 467 fs
lifetime. Experimental searches must therefore cover a range of possible
lifetimes.

We consider two possible scenarios for the decay of Pentaquark baryons. In the
first, Pentaquarks are quasistable hadrons which can decay only via weak
interaction, with lifetimes of the same order as other charmed hadrons.
Possible decay modes of Pentaquarks in this scenario are $P^0~\rightarrow~ \phi
p \pi^-,~ \Lambda K^+\pi^-;~P^-~\rightarrow p \phi \pi^- \pi^-,~\Lambda K^+
\pi^- \pi^-, ~\Sigma^- K^+ \pi^-$; as well as those described in the previous
paragraph. Such decays can be directly observed in precision vertex detectors,
which are now standard devices in experiments with charmed or beauty particles.
The use of vertex detectors greatly reduces the combinatorial background in the
corresponding effective mass spectra, and makes it possible to search for
exotic $P$ baryon production in a wide range of the $X_f$ observable. Searches
for the tetraquark $\tilde{F}_s$ ($cs\bar{u}\bar{d}$) could also look for weak
decays \cite {lip,lg}, such as $\tilde{F}_s \rightarrow K^- K^+ \pi^- \pi^+$.

In the second scenario, the Pentaquark baryons have large enough masses and are
resonant states, which strongly decay with emission of secondary charmed
particles (for example, $P^0~\rightarrow~ pD^-_s$). The search for such
strongly decaying exotic hadrons must also use vertex detectors for detection
of the $D_s^-$. For the identification of Pentaquarks, only the study of
effective mass spectra of secondary decay products can be used, for which there
is a large combinatorial background. It is well known \cite{lgl,lg} that the
combinatorial background is significantly reduced in the fragmentation region
(at $X_f~>~0.5 - 0.6$), and such kinematics is therefore strongly desirable for
the resonance Pentaquark searches. This experimental task is more challenging
and is crucially dependent on the Pentaquark decay width. Only narrow states
with $\Gamma~\leq~100$ MeV have a good chance to be separated from the
background. Narrow states may arise for a variety of reasons not necessarily
related to  exotic properties, as when the phase space for decay is small. For
example, the $\Lambda$(1405) is 80 MeV above the $\Sigma \pi$ threshold and has
a width of 50 MeV. The D$^*$(2010) and the $\Lambda^*_c$(2625) are about 40 MeV
above their respective thresholds and their widths are less than 2 MeV. Another
important possibility is a narrow width that may arise from the complicated
internal color structure of an exotic hadron, and by a quark rearrangment
mechanism in the decay processes for multiquark exotic object, leading to a
colorless final state \cite {A,lg,cl}. Another possible cause for a narrow
state may be the reduced effective phase space as a result of OZI suppression
of some decays for  an exotic state with hidden strangeness or charm \cite
{lgl}, as will be discussed in Section 5.

\section*{4.~~Experimental Pentaquark Search}

An experimental program to search for the Pentaquark should include:\\
(1) Reactions likely to produce the Pentaquark, complemented by an estimate
of the production cross section.\\
(2) Experimental signatures that allow identification of the Pentaquark.\\
(3) Experiments in which the backgrounds are minimized. \\
These points will be further discussed in the
following subsections.

\subsection*{4.1~~Experimental Considerations}

All charm experiments require vertex detectors consisting of many planes of
silicon micro-strips with thousands of channels. Fermilab E791 \cite
{791,jl791,e791} used 23 such planes. Some of the planes are upstream of the
target for beam  tracking. These detectors allow a high efficiency and high
resolution for reconstruction of both primary (production) vertex and secondary
(decay) vertex. The position resolution of the vertex detectors is typically
better than 300 microns in the beam direction. By measuring the yield of a
particle as a function of the separation between the two vertices, the lifetime
of the particle is obtained. This is possible as long as  the lifetime  is not
so short, such that the separation of vertices becomes ambiguous. Other major
components in charm experiments are several magnetic spectrometers with track
detectors for track reconstruction and for momentum analysis, Cherenkov
counters for particle identification, and electromagnetic and hadronic
calorimeters. Muon detectors and TRD detectors for electron separation are
included for studies of leptonic decays. The invariant mass resolution for
typical charm masses in such spectrometers is about 10 MeV. Different
spectrometers are sensitive to different regions of Feynman $X_f$ values.

In hadronic production, the charm states produced are preponderantly charm
mesons at low $X_f$. The triggers for such experiments vary. In E791, the
requirement was to ensure an interaction in the target (using signals from
various scintillators) and a transverse energy ($E_{t}$) larger than some
threshold \cite {791,jl791,e791}. The rest of the charm selection was done
off-line. Future experiments are planned to obtain higher yields of charmed
hadrons. Increased charm sensitivity can be achieved as in E781 \cite {russ} by
using higher integrated beam intensities, and higher efficiency detector
systems. E781 also will use a trigger condition that identifies a secondary
vertex, and also requires positive particles with momentum greater than 15
GeV/c. This should enhance the high-$X_f$ acceptance ($X_f>0.1$), and give
higher quality events. A good charm trigger \cite {russ} can produce an
enriched sample of such high $X_f$ charm baryons with improved reconstruction
probability because of kinematic focusing and lessened multiple scattering with
a significantly lower number of events written to tape or disk. CHARM2000
experiments will also require charm enhancement triggers \cite {jeff}. The
present E791 \cite {e791} and future E781 \cite {russ} and CHARM2000
experiments \cite {dk,paul,dct} complement each other in their emphasis on
different $X_f$ regions, incident particle types, statistics and time
schedules.

High quality particle identification (PID) for the largest possible energy
range of the outgoing particles is important for reducing backgrounds
associated with incorrect identification of tracks. In E791, two threshold
Cerenkov detectors were used for this purpose. In E781, this will be available
via ring imaging Cerenkov (RICH) and transition radiation detector (TRD) PID
systems. These and other experimental techniques to reduce backgrounds are
described in more detail in \cite{kk}.

\subsection*{4.2~~Pentaquark Production Mechanisms}

 We consider possible mechanisms for P formation. For the central
hadron-nucleus charm production at several hundred GeV/c, the elementary
process is often associated with $q\bar{q} \rightarrow c \bar{c}$ or $gg
\rightarrow c \bar{c}$ transitions. The produced charmed quarks propagate and
form mini-jets as they lose energy. Hadronization associated with each jet
proceeds inside the nucleus, and to some extent also outside the nucleus;
depending on the transverse momentum of the jet. The propagating charmed quarks
may lose energy via gluon bremsstrahlung or through color tube formation  in a
string model, or by other mechanisms, as discussed in ref. \cite {nied} and
references therein. One may form a meson, baryon, or Pentaquark, according to
the probability for the charmed quarks to join together with appropriate quarks
and antiquarks in the developing color field. One can estimate Pentaquark
production cross sections via one-step and also two-step hadronization. All
such estimates are very rough. Our aim is to account for major ingredients in
estimating the cross section, and to give a conservative range of values. For
one-step hadronization, the $\bar{c}$ joins directly to the other quarks. The
one-step is the usual mechanism for meson and baryon formation. For two-step,
the first involves meson and baryon hadronization, while the second involves
meson-baryon coalescence.

We first consider estimates for the central production cross section assuming a
meson-baryon coalescence mechanism, expected to be the main mechanism for
production through the long-range (deuteron-like) component of the Pentaquark
wave function. We make a crude estimate relative to the D$_{s}^-$, an
anticharmed-strange meson ($\bar{c}s$). The weakly bound P (deuteron type
structure) can be produced for example by coalescence of a proton or a neutron
with a D$_{s}^-$, analogous to the production of a deuteron by coalescence of a
neutron and a proton. The data \cite {dbar} give roughly 10$^{-3}$ for the
$\sigma(d)/\sigma(p)$ production ratio. This ratio can also be applied to
$\sigma(P)/\sigma(D_{s}^-)$ production. The reason is that in both cases, the
same mass (nucleon mass) is added to the reference particle (proton or
D$_{s}^-$), in order to form a weakly bound deuteron-like state. This ratio is
very sensitive to the pentaquark binding energy, and may be substantally
different for a larger value.

We now consider the one-step hadronization of a Pentaquark, expected to be the
main mechanism for the production through the short-range component of the
Pentaquark wave function. We rely here on an empirical formula which reasonably
describes the production cross section of a mass M hadron in central
collisions. The transverse momentum distribution at not too large p$_t$ follows
a form given as \cite {hag}:
\begin{equation}
d\sigma/dp_t^2 \sim exp(-C\sqrt{M^2+p_t^2}),
\end{equation}
where C is roughly a universal constant $\sim$ 5 - 6 (GeV)$^{-1}$. The
exponential (Boltzmann) dependence on the transverse energy $E_t =
\sqrt{M^2+p_t^2}$ has inspired speculation that particle production is thermal,
at a temperature C$^{-1}$ $\sim$ 160 ~MeV \cite {hag}. We assume that this
equation is applicable to Pentaquark production. To illustrate the universality
of C, we evaluate it for a few cases. For $\Lambda_c$ and $\Xi^0$, empirical
fits to data give exp(-$bp_t^2$), with b=1.1 GeV$^{-2} $and b=2.0 GeV$^{-2}$,
respectively \cite{wa89,rot}. With C $\approx 2 \cdot b \cdot M$, this
corresponds to C $\sim$ ~5.0 GeV$^{-1}$ for $\Lambda_c$, and C $\sim$ ~5.3
GeV$^{-1}$ for $\Xi^0$. For inclusive pion production, experiment gives
exp(-$bp_t$) with b =~6 GeV$^{-1}$ \cite {pi6}; and C $\sim$ b, since the pion
mass is small. Therefore, C= 5-6 GeV$^{-1}$ is valid for $\Lambda_c$, $\Xi^0$
hyperon, and pion production. After integrating over p$_t^2$, we estimate the
ratio as:
\begin{equation}
\sigma(P)/\sigma(D_{s}^-) \sim exp[-5[M(P)-M(D_{s}^-)]] \sim 10^{-2}.
\end{equation}
\noindent
In applying Eq. 5 to Pentaquark production, we assume that the suppression of
cross section for the heavy P as compared to the light D$_{s}^-$ is due only to
the increased mass of P. The formula ignores dynamical considerations, such as
the particular one-step or two-step hadronization processes, or the size of the
P. It also does not account for threshold effects. Consequently, there may be
large uncertainties in its application.

Both reaction mechanisms described above can contribute to the production cross
section, which is estimated in the range of $\sigma(P)/\sigma(D_{s}^-) \sim
10^{-3} - 10^{-2}$. In actual measurements, the product $\sigma$ $\cdot$ B for
a particular decay mode is measured, where B is the branching ratio for that
mode. If a Pentaquark peak is not observed, assumptions on the values of B and
on the P lifetime may be necessary in order to set limits to the Pentaquark
production cross section.

Another approach for the $\sigma(P)$ cross section evaluation is based on a
comparison of Pentaquark and charmed-strange baryon $\Xi_c^0$ (csd) production
reactions. This method was previously applied in \cite{791,jl791,lg}. For
example, with a high energy hyperon beam, one may compare P$^0$ production
$(uud\bar{c}s)$  to $\Xi^0_c$ $(csd)$ baryon production. Processes with charm
baryon production involve $c\bar c$ pair creation, followed by $c$ or $\bar c$
hadronization in the final states. There should therefore not be much
difference for $c$ or $\bar c$ fusion, so that one may compare $P^0$ to
$\Xi^0_c$ production. The production of $P^0$ is also different from $\Xi^0_c$
due to the fusion  of an additional uu quark pair. This would cause a reduction
factor R for the cross sections:
\begin{equation}
\sigma_{\Sigma}(P^0)~=~R
\sigma_{\Sigma}(\Xi^0_c)~\geq~2.5 \times 10^{-3}
\sigma_{\Sigma}(\Xi^0_c).
\end{equation}
This factor $R~\geq~2.5\times 10^{-3}$ can be estimated from the data on other
processes with additional quark fusion in the particles under study. For
example, the relative yields $\bar d/\bar p~\simeq~10^{-4}, \bar{^3 He}/\bar
p~\simeq~10^{-8}$ \cite{nied,dbar,lmk} give information on the fusion
probability of 3 and 6 additional quarks. It is then possible to obtain an
average reduction factor  $\sim~5 \times 10^{-2}$ per fusion for each
additional quark, yielding the value $R~>~2.5\times 10^{-3}$ in Eq. 6. The
inequality arises because antinuclei are loosely bound systems whose yields
implicitly include this factor.

\subsection*{4.3~~Pentaquark Decay Signatures}

(1) Mass, Width  and Decay Modes

   Searches for the Pentaquark are easiest via modes having all final decay
particles charged. With all charged particles detected, the invariant mass of
the system can be determined with high resolution. One signature of the
Pentaquark is a peak in the invariant mass spectrum somewhat lower than 2907
MeV if the system is bound; and above if it is a resonance. The position of the
peak should be the same for several decay modes. It's width should be
determined by the experimental resolution if it is bound, and broader if it is
a resonance.

The selection of the decay modes to be studied is made primarily by considering
detection efficiency and expected branching ratios. Since the $D_{s}^-p$ system
is the lightest, it is expected to be preferred from phase space arguments.
Also, two of it's decay modes have four charged particles in the final state
(e.g. $\phi\pi^-p$, $\phi \rightarrow K^+ K^-$; ${K^{*0}}K^-p$, ${K^{*0}}
\rightarrow K^+ \pi^-$). This signature was implemented in E791 \cite
{791,jl791}. First, two distinct vertices were identified, a production vertex
and a decay vertex. From the decay vertex, four tracks were identified and
associated with $K^+K^-\pi^-p$. By reconstructing the invariant mass of the
$K^+K^-$ pair, only $\phi$ mass events were accepted and the invariant mass of
all four particles was reconstructed. A peak in the resulting spectrum will be
one of the identifying characteristics of the Pentaquark.

(2) One General Signature - A Spectator Baryon:

We first note a striking signature for Pentaquark decay which may be useful for
discrimination against background. This signature is predicted by both of two
very different Pentaquark models: (1) a loosely-bound $D_{s}^-p$ deuteron-like
state and (2) a strongly-bound five-quark state. Both models predict decay
modes into a baryon and two or more mesons, in which the three quarks in the
baryon are spectators in the decay process and remain in the final state with a
low momentum which is just the fermi momentum of the initial bound state.

That the baryon is a spectator is obvious in the deuteron model, in which
the decay is described as an off-shell $D_{s}^-$ decaying with a nucleon
spectator. In the five-quark model, a similar situation arises in the
commonly used spectator model with factorization. Here, the charmed
antiquark decays into a strange antiquark by emission of a $W^-$ which then
creates a quark-antiquark, which hadronizes into mesons. The strange
antiquark combines with one of the four spectator quarks to form one or
more mesons, while the three remaining spectator quarks combine into a
baryon.

In both cases, it seems that the final state should show a low-momentum
baryon in the center-of-mass system of the Pentaquark and the invariant
mass spectrum of the remaining mesons peaked at the high end near the
kinematic limit. Thus in the particular cases of the $p \phi \pi^-$,
$K^{*o}K^-p$ and $\Lambda K^+ \pi^-$ decay modes, the $\phi \pi^-$, $K^{*o}
K^-$ and $K^+ \pi^-$ invariant mass distributions respectively should show
this peaking near the kinematic limit.

Note that in the particular case of the $p \phi \pi^-$ decay mode, a low
momentum proton in the center of mass system means that the $\pi^-$ and
$\phi$ are back to back with the same momentum and therefore that the pion
carries off most of the available energy. Thus one might reduce background
with a cut that eliminates all pions with low momentum in the center of
mass.

(3)  Some Model-Dependent Branching Ratio Predictions:

The $\phi\pi^-p$ decay mode is the most convenient for a search, since the
$\phi$ signal is so striking. We now examine the lowest order predictions from
the two extreme models for the branching ratios of other modes relative to
$\phi\pi^-p$.

In experiments sensitive only to charged particles the $\phi\pi^-p$ decay mode
is observed in the four-prong final state $ K^+ K^- \pi^- p$. The $K^{*o}K^-p$
decay mode is also observable in this same four prong final state. The $K^{*o}
K^-p$ decay mode arises naturally in the deuteron model, since the $K^{*o}
K^-$ decay is observed for $D_{s}^-$ decays with a comparable branching ratio
to
$\phi \pi^-$. In this model, the ratio of the two decays is predicted from
observed $D_{s}^-$ decay branching ratios with phase space corrections.
However,
the $K^{*o} K^-p$ decay mode does not occur in the five quark spectator model,
where the spectator strange quark can only combine with the $\bar s$ produced
by the charm decay to make a $\phi$ or with two spectator nonstrange quarks to
make a hyperon. Comparing the two decays thus tests the decay model.

The $K\pi\Lambda$ and $K^*\pi\Lambda$ decay modes arise naturally in the five
quark spectator model, or in a moderately bound D$\Lambda$ Pentaquark. However,
they should not be expected in a very weakly bound deuteron model with mainly a
$D_{s}^-$p structure. In that case, the $D_{s}^-$ decays into mesons containing
one strange quark-antiquark pair and the baryon spectator has no strangeness.

(4) Angular Momentum Constraints and Angular Distributions
for P Decays:

We can give a model-independent prediction. The Pentquark has spin 1/2 and this
total angular momentum is conserved in the decay. Since the production process
is a strong interaction which conserves parity, the Pentaquark will not be
produced with longitudinal polarization. Its polarization in the beam direction
must also vanish. Therefore, the angular distribution in the center-of-mass
system of the Pentaquark must be isotropic for the momentum of any final state
particle in any decay mode with respect to either the incident beam direction
or the direction of the total momentum of the Pentaquark. The background does
not necessarily have these constraints.

We also give a model-dependent prediction. We first consider the deuteron
model. The $D_{s}^-$ has spin zero, and spin is preserved in the decay. Thus,
in the center of mass frame of all the $D_{s}^-$ decay products, the angle
between the proton momentum and the momentum of any particle emitted in the
$D_{s}^-$ decay must have an isotropic angular distribution.

A further prediction is obtainable for the case of a vector-pseudoscalar
decay mode of the $D_{s}^-$; e.g. $\phi\pi^-$ or $K^{*0} K^-$. The vector meson
must be emitted with zero helicity in the rest frame of the $D_{s}^-$. The zero
helicity can be seen in the  $\phi \pi$ decay by measuring the angle $\theta_{K
\pi}$ between the kaon momenta in the $\phi$ rest frame and the pion momentum.
The prediction is to have a $\cos^2 \theta_{K \pi}$ distribution. By contrast,
the five-quark model for the Pentaquark favors helicity one over helicity zero
for the vector meson by just the 2:1 ratio needed to give an isotropic
distribution in $\theta_{K \pi}$. Here again the background does not
necessarily have these constraints.

\subsection*{4.4~~Pentaquark Expected Yield}

We proceed with count rate estimations for the expected yields of Pentaquark
baryons. Given the need to search for both quasistable and resonant Pentaquark
baryons in different regions of $X_f$, we give production cross sections for
$X_f~>~0$ and in the fragmentation region $X_f~>~0.5$, where one expects an
improved signal to background ratio. We use several production models for
Pentaquarks. The different predictions of these models reflect the
uncertainties in our expectations. We also assume that quasistable $P^0$
baryons would be reconstructed in selected visible weak (w) decay modes with a
combined effective branching ratio B$_w$:
\begin{eqnarray}
\nonumber B_{w}=~B[P^0\rightarrow p \phi \pi^- + p \phi  \pi^- \pi^+ \pi^- + p
K^{0*} K^-] \\
B_{w}=~\simeq~0.05~; ~based~ on~ \phi \rightarrow K^+ K^-
and ~ K^{0*} \rightarrow K^+ \pi^-.
\end{eqnarray}
In practice, one looks for the visible (charged-particle) decay modes, such as:
$[D_{s}^-p] \rightarrow \phi \pi^- p \rightarrow K^+ K^- \pi^- p$ (B=1.8\%),
$[D_{s}^-p] \rightarrow \phi \pi^- \pi^- \pi^+ p \rightarrow K^+ K^- \pi^-
\pi^- \pi^+ p$ (B=0.9\%), and $[D_{s}^-p] \rightarrow K^{*0}K^-p \rightarrow
\pi^- K^+ K^-p$ (B=2.1\%). The total branching ratio of these last three decay
modes is 4.8\%. Here we assume a $[D^-_s p]$ bound state model for the
quasistable P pentaquark. Our estimate $B_{w}= \simeq~0.05$ is somewhat lower
than the total visible (charged particle) decay modes of a virtual D$^-_s$ for
$P^0 \rightarrow (p D^-_s)_{virtual}$ dissociation, where $B(D^-_s)(visible)
\approx 0.08$ \cite {pdg}. The 0.08  value includes possible non-resonant
charged particle decays, such as $D_s^- \rightarrow p K^+ K^- \pi^-$ or $D_s^-
\rightarrow p \pi^+ \pi^- \pi^-$. Our objective is a search that will have
improved signal to background, by concentrating on final states having
resonances, such as the narrow $\phi$ or the K$^*$(890).

For strongly-decaying P baryon resonances, we assume the same production cross
sections as for a quasi-stable Pentaquark. A reasonably small decay width can
be obtained in principle in the model with direct five-quark configurations
(color octet or sextet bonds, as described in Section 3 and Ref. \cite {A}).
For detection of these baryons, one may hunt for the decay mode $P^0
\rightarrow p D^-_s$, assigned here a 100\% branching ratio for this resonance
possibility. This value is clearly an upper limit, and may be reduced by other
strong decay channels. We for example do not take into account other possible
totally reconstructable decay modes, such as $P^0 \rightarrow p D^-_s(visible)
\pi^+\pi^-$, $P^0 \rightarrow \Delta^0  D^-(visible) K^+$. Thus, we estimate
for strongly decaying pentaquarks, the branching ratio:
\begin{equation}
B_s[P^0](visible)~=~B_{s}=B[P^0 ~\rightarrow~pD^-_s]
\cdot B[D^-_{s} visible]~\simeq~0.05.
\end{equation}
One may also search for $P^-$ Pentaquark strong decays $P^- \rightarrow
D^-\Lambda;~\bar{D^0} \Sigma^-$ (resonances).

Both weak and strong decay modes coming from the $D_{s}^-p$ and the
$\overline{D^{0}}\Lambda$ components of the P are currently being studied in
E791, where the data were taken with a 500 GeV $\pi^-$ beam. Analysis of a part
of the E791 data already yielded a preliminary upper limit at 90\% confidence
level:
\begin{equation}
\frac{\sigma(P^0)\cdot B(P^0 \rightarrow \phi\pi p)} {\sigma(D_s^-)\cdot
B(D_s\rightarrow\phi\pi)} < 2.6\%,
\end{equation}
for quasistable Pentaquark production (not including systematic uncertainties)
\cite{sdpf}. In the analysis, it was assumed that the Pentaquark has the same
lifetime as a real $D^-_s$ and a mass of 2.75 GeV. It was assumed that its
production characteristics are the same as other charmed baryons. This limit
was based on a part of the data and measured $D^-_s$ yield. With the full data
sample and more decay modes analyzed, several Pentaquarks may be observed if
the cross section is in the range estimated in the previous section, or else
the limit may be further lowered.

For the planned E781 and CHARM2000, when both use baryon beams, we rely on
previous measurements done with similar beams. The $\Sigma^-$ hyperon beam
should be good tool for the search for strange-charmed(anticharmed) open exotic
hadrons, such as the Pentaquark baryons $P^0$ and $P^-$, or tetraquark meson
$\widetilde{F}_s$. Hyperon beams are the purest high energy beams containing
strange valence quarks. Thus, the inclusive reactions:
\begin{eqnarray}
\nonumber \Sigma^-(dds)+N & \rightarrow & P^0(\bar cuuds)+X \\
\nonumber & \rightarrow & P^-(\bar cudds)+X \\
& \rightarrow & \widetilde{F}^0_s(cs \bar u \bar d)+X
\end{eqnarray}
should be favorable for the production of strange-charmed hadrons at large
X$_f$, since they benefit from strange quark sharing between primary and
secondary particles.

We use different methods to estimate the expected cross sections for Pentaquark
production in $\Sigma^-N$ interactions for the hyperon beam of Fermilab with
momentum 650 GeV/c (E781). First, following section 4.2, we take
$\sigma_{\Sigma^-}(P^0)/\sigma_{\Sigma^-}(D^-_s)~\simeq~ 10^{-2}-10^{-3}$. The
$\sigma_{\Sigma^-}(D^-_s)$ cross section is estimated two ways. A neutron beam
measurement with $E_n~\simeq~600$ GeV \cite {ds} gave
$\sigma_{n}(D^-_s)~\cdot~B(D^-_s~\rightarrow~\phi \pi^-)~\simeq ~0.76\mu b/N$
for $0.05~ <~ X_f~ <~ 0.35$. From the quark structure of $\Sigma^-$ and the
neutron, one may expect that $\sigma_{\Sigma}(D^-_s)$ is greater than
$\sigma_{n}(D^-_s)$. Assuming conservatively that they are equal,  and assuming
also the $X_f$ dependence $d\sigma(D^-_s)/dX_f\sim (1-X_f)^5$ for the baryon
primaries (see \cite{wa89}), one obtains
$\sigma_{\Sigma}(D^-_s)|_{X_f>0}~\simeq~40~ \mu b/N$. This is an exceptionally
large value,  considering that the entire hadronic (cq) production cross
section is expected \cite {alm} to be near 20-30 $\mu$b/N. A more conservative
estimate of this cross section is based on the assumption
$\sigma_{\Sigma}(D^-_s)~\simeq~ \sigma_{p}(D^-)~\simeq~\sigma_{p}(D^+)$ from
the quark structure of the projectiles and produced mesons.  The data of the
EGS experiment for $X_f>0$  with a $P_p=400$ GeV/c momentum  proton beam give
$\sigma_{p}(D^-)~\simeq~\sigma_{p}(D^+)~\simeq~3~ \mu b/N$ \cite {ab}. Based on
the energy dependence of charm production data \cite {appel}, extrapolation to
650 GeV/c yields for E781 the estimate
$\sigma_{\Sigma}(D^-_s)|_{X_f>0}~\simeq~5~ \mu b/N$. This smaller value is used
for the estimate of $P^0$ exotic baryon yields in E781 (see Table 1). With
$\sigma(P^0)/\sigma(D^-_s)\simeq 10^{-2}-10^{-3}$, one obtains $\sigma(P^0)
\approx 50-5~ nb/N$.

There are no direct data for the $\Xi ^0_c$ production with baryon beams, and
only very poor data for $\Xi^+_c$ (csu) production \cite{Cq}. The recent WA89
experiment  used a $\Sigma^-$ hyperon beam at CERN, with momentum
$P_{\Sigma^-}$=330 GeV/c. The cross section measured was:
$\sigma_{\Sigma^-}(\Lambda^+_c)|_{X_f>0.2} = (9.3 \pm 4.3 \pm 2.5)~ \mu b/N$,
with $d\sigma_{\Sigma}/dX_f~\sim~(1-X_f)^4$ \cite{wa89}. Extrapolating to
$X_f~>~0$ (factor $\sim$ 3) and to $P_{\Sigma^-}$=650 GeV/c
(factor$~\sim~1.5$), one obtains
$\sigma_{\Sigma^-}(\Lambda^+_c)|_{X_f>0}~\simeq~(30 \pm 14 \pm 8) ~\mu b/N$ for
E781, where the uncertainties are only from scaling those of Ref. \cite {wa89}.
The value 30 $\mu b/N$ again seems unreasonably large, which may be due to the
extrapolation procedure or to the experimental uncertainties in
$\sigma_{\Sigma^-}(\Lambda^+_c)|_{X_f>0.2}$. Thus, we use the more conservative
estimate $\sigma_{\Sigma^-}(\Lambda^+_c)|_{X_f>0}~\sim~5~\mu b/N$ for
$P_{\Sigma^-}$=650~Gev/c, close to the minimum value given by the error bars.
{}From the quark composition of charmed baryons, it is reasonable to conclude
that $\sigma_{\Sigma}(\Xi^0_c)~\geq~\sigma_{\Sigma}(\Xi^+_c)~
or~\sigma_{\Sigma}(\Lambda^+_c) ~\geq~5~\mu b/N$; and to obtain from Eq. (6)
the estimate:
\begin{equation}
\sigma_{\Sigma}(P^0)|_{X_f>0}~\geq~13~nb/N.
\end{equation}
This value gives Pentaquark yields in E781 at about 25\% of the upper values
shown in Table~1.

We consider also the expected E781 efficiency for Pentaquark detection, by
comparison to estimated \cite {russ} efficiencies for cqq decays. These
include a tracking efficiency of 96\% per track, a trigger efficiency averaged
over $X_f$ of roughly 18\%, and a signal reconstruction efficiency of roughly
50\%. The  E781 Monte Carlo simulations \cite {russ} gave an average global
efficiency of $\sim$ 8\%, by considering the $\sim$ 200 ~fs lifetime decay
$\Lambda^+_c \rightarrow p K^- \pi^+$, and the $\sim$ 350 ~fs lifetime decay
$\Xi^+_c \rightarrow \Xi^- \pi^+ \pi^-$. The charm baryons were assumed \cite
{russ} to be produced with a cross section of the form
$d\sigma/dX_f=(1-X_f)^{4.2}$, an assumption which is built into the estimation
of the trigger efficiency. For heavier Pentaquark production, it is likely that
this distribution would shift to lower X$_f$ (corresponding to an exponent
greater than 4.2), and this would also reduce the efficiency. The value 8\% is
for reconstruction of the relatively strong signals from cqq charm baryon
decays. The reconstruction efficiency may be lower for Pentaquark events. For
the weaker Pentaquark signal, tighter analysis cuts with resulting lower
efficiencies may be required in order to achieve the optimum signal to noise
ratio. For lifetimes smaller than about 60 ~fs, which is possible for the
Pentaquark, the trigger efficiency would also be significantly reduced \cite
{russ}. Considering all the unknown variables, the final experimental
statistics may then be significantly lower than the upper value estimated here,
using the 8\% global efficiency. From Table 1, one sees that for quasistable
Pentaquark baryons, the maximum expected statistics in E781 is 280-2800 events.
This may be adequate for their observation, if they really exist.

{}From $d\sigma/dX_f \sim (1-X_f)^4$ for charmed baryons \cite {wa89}, we
estimate the Pentaquark production cross section in the $X_f~ >~ 0.5$
fragmentation region (quasi-stable or resonant) as $\sigma_d \sim 0.03 \cdot
\sigma(X_f > 0)$; as given in Table 1. The fragmentation region should be most
effective for reducing the combinatorial background in both quasi-stable and
resonance Pentaquark searches. In the latter case, one studies a strong decay
into $D_{s}^-p$, if the P is a short-lived resonance with mass
$M(P^0)~>~M(D_s^-)~+~M(N)$. For this strong decay, the proton and D$^-_{s}$
come from primary vertex, and the D$^-_{s}$ decay forms the secondary vertex.
The expected maximum numbers of events in the fragmentation region is quite
limited (from 28 to 280 events, from Table 1), and there are moderate chances
for P$^0$  observation as a peak in the mass spectrum of $M(pD^-_s)$.

 It is possible that different mechanisms for charm production contribute in
different $X_f$ regions. For example, there is evidence for leading production
of charmed hadrons in WA89 and FNAL E769 \cite{bal} and E791 \cite {e791l}. One
can also consider a diffractive mechanism for Pentaquark production. Brodsky
and Vogt \cite {bro,lead} suggested that there may be significant intrinsic
charm (IC) $c\bar{c}$ components in hadron wave functions. The Hoffmann and
Moore analysis \cite {hm} of charm production in deep inelastic electron
scattering yields 0.3\% IC probability in the proton. A recent reanalysis of
the EMC charm production data was carried out by Harris, Smith, and Vogt \cite
{hsv}. Their improved analysis found that an IC component is still needed to
fit the EMC data, with a value indicated for the proton of (1.0 $\pm$ 0.6)\%.
The most probable IC state occurs when the constituents have the smallest
invariant mass.  In the rest system, this happens when the constituents are
relatively at rest.  In a boosted frame, this configuration corresponds to all
constituents having the same velocity and rapidity \cite {bhmt,intc}. When a IC
state is freed in a hadronic collision, the charm quarks should have
approximately the same velocity as the valence quarks. They can then easily
coalesce into charmed hadrons \cite {lead,bgs,mbs,vdw} and produce leading
particle correlations at large $X_f$. The IC component in an incident
$\Sigma^+$ or proton can then lead to large $X_f$  $P^0$ production. One
may expect that P$^0$ production will be predominantly central for reaction
mechanisms other than IC. Intrinsic charm Pentaquark production, with its
expected high X$_f$ distribution, would therefore be especially attractive.

For E781 and CHARM2000 one can study the diffractive pair production reactions
(\cite{mm}, see also \cite{lg}) $\Sigma^-+N \rightarrow (P^-D^0)+N$ and $p+N
\rightarrow (P^0D^+_s)+N$ (or even the coherent reactions of these types on
nuclei), with possible $D^0$, $D_{s}^+$ tag or without such tag. For the
diffractive pair production cross section, one can compare to the diffractive
cross section for the reaction $p + N \rightarrow (\Lambda K^+) + N$ at 70 GeV,
which is more than 1 $\mu b$ after subtraction of isobar contributions
\cite{lg}. The ratio of diffractive cross sections $\sigma_d$ was estimated
\cite{lg}:
\begin{equation}
\sigma_d(P^0 D_{s}^+)[600 GeV]/\sigma_d(\Lambda K^+)[70 GeV] \sim
(m_s/m_c)^2 \cdot R  \sim 2. \times 10^{-4}.
\end{equation}
Here the reduction factor R that accounts for the fusion in the P of an extra
two quarks is R=$2.5 \times 10^{-3}$ from Eq. 6. The factor $(m_s/m_c)^2$
accounts for the relative probability to produce charmed quark versus strange
quark pairs. From the ratio of constituent quark masses \cite {cq},
$(m_s/m_c)^2 \sim 8. \times 10^{-2}$. In Eq. 12, we do not explicitly show a
kinematic factor, since its value \cite {lgl,lg} is close to unity for the
reactions shown. This factor accounts for the energy and mass dependence of the
cross sections, for diffractive-like production of different final state masses
with different beam energies. The increased Pentaquark cross section expected
from energy extrapolation is offset by a reduction for making the heavier mass
Pentaquark final state. We then find:
\begin{equation}
\sigma_d(P^0 D_{s}^+)[600 GeV] \sim 1~ \mu b \times 2. \times 10^{-4}
\sim 0.2 ~nb.
\end{equation}
We assume conservatively that $B_w$=0.05 for such diffractive production
searches, although with the smaller backgrounds expected at high $X_f$, one
could possibly search for all visible decays ($B_w$=0.08). From the projected
Pentaquark charm sensitivity shown in Table 1 for E781 (and roughly ten times
higher for CERN CHEOPS \cite {paul}), it may be possible to observe the
diffractive production process of Pentaquarks at the level of several hundred
events. Future  experiments in the spirit of the CHARM2000  workshop, with
higher charm sensitivity than E781, should have improved chances to observed
Pentaquarks. \\

\section*{5.~~Heavy Baryons with Hidden Charm}

 In recent years, several candidates were reported for baryon states  with
unusual properties (narrow decay widths, large branching ratios for the decays
with strange particles). There are candidates for cryptoexotic baryons with
hidden strangeness $B_{\phi}=\mid$~$qqqs\bar{s}\!>$, were $q=u$ or $d$ quarks
(see \cite{lgl,ks,ccc} and references therein). Further searches  for
nonstrange and cryptoexotic baryons with hidden strangeness are planned at IHEP
(70 GeV proton beam) with the SPHINX spectrometer \cite{ccc}, and in FNAL E781
\cite {111}. Although the existence of such  a $B_{\phi}$ baryon is not yet
confirmed, this suggestion raise the question of the possible existence of
heavy cryptoexotic baryons with hidden charm $B_{\psi}=\mid$~$qqqc\bar{c}\!>$.
These exotic Pentaquark baryons can either be direct five-quark states or
$\phi-N$ and $\Psi-N$ bound states (resonances). The latter may form at low
relative velocities of the meson and baryon consituents, because of the strong
attractive QCD Van der Waals interaction \cite {vdw}. The intrinsic charm which
was discussed in Section 4.4 may also be relevant for the production of the
B$_{\Psi}$ baryon. If M$(B_{\psi})<$M$(\eta_{c})+$M($p$) $\simeq$ 3.9~GeV, the
$B_{\psi}$ decays would be OZI suppressed and the width of this cryptoexotic
baryon would be quite narrow. If M$(B_{\psi})>$4.3~GeV, there would be OZI
allowed decays $B_{\psi}^+ \rightarrow p + J/\psi; \Lambda^+_C + D^0$, etc.
Because of a complicated internal color structure of this baryon (see
Introduction), one can expect a narrow decay width ($\leq$~100~MeV).

To search for such $B_{\psi}$ states, it was proposed \cite {lglm} to use the
diffractive production reaction $p+N \rightarrow B_{\psi}^+ + N$ with possible
decays of $B_{\psi}$ baryons $B_{\psi}^+ \rightarrow p + (J/\psi)_{virt}
\rightarrow p+ (l^+ l^-)$ or $B_{\psi} \rightarrow p+(\eta_c)_{virt}
\rightarrow p + (K^+K^-\pi^+\pi^-;2\pi^+2\pi^-; K\bar{K}\pi;\eta\pi\pi)$. For
these processes, $\sigma  B  \epsilon$ was estimated as 0.2~ - ~0.5 ~nb/N
\cite{lglm}. For an appropriately designed experiment, with a charm sensitivity
close to 10$^4$ events/(nb/N) of effective cross section, it should be possible
to observe several thousand such events.

\section*{6.~~Conclusions}

We described the expected properties of Pentaquarks. Possibilities for
enhancing the signal over background in Pentaquark searches were investigated.
General model-independent predictions were presented as well as those from two
models: a loosely bound $D_{s}^-N$ ``deuteron" and a strongly-bound five-quark
model. While the current E791 may have marginal sensitivity, future experiments
with more than $10^6$ reconstructed charmed baryon events should have enough
sensitivity to determine whether or not the Pentaquark exists.

\section*{7.~~Acknowledgements}

Thanks are due to J. Appel, S. Brodsky, P. Cooper, L. Frankfurt, S. Gavin, D.
Kaplan, K. Konigsmann, M. A. Kubantsev, S. Kwan, J. Lach, J. Lichtenstadt, S.
May-Tal Beck, B. Muller, S. Nussinov, S. Paul, B. Povh, J. Russ, I. I.
Strakovsky, B. Svititsky, and R. Vogt  for stimulating discussions. This work
was supported in part by the U.S.-Israel Binational Science Foundation,
(B.S.F.) Jerusalem, Israel, by grant No. I-0304-120-07/93 from the
German-Israeli Foundation for Scientific Research and Development, and by
Russian Ministry of Science, Moscow, Russia. \\

E-mail addresses of authors are:\\
murraym@tauphy.tau.ac.il, ashery@tauphy.tau.ac.il,\\
lgl@mx.ihep.su, ftlipkin@weizmann.weizmann.ac.il

\newcommand{\quasistable}{\vbox{\hbox{\strut quasi-} \hbox{stable}}}
\newcommand{\shortlived}{\vbox{\hbox{\strut short-} \hbox{lived}}}
\begin{table}[tphb]
\caption{Projected Pentaquark Yields}
\scriptsize
\begin{tabular}{|c|c|c|} \hline
\multicolumn{2}{|c|}{Model for estimation} &
$\sigma(P^0)/\sigma(D^-_s)\simeq 10^{-2}-10^{-3}$ \\[1mm]
\multicolumn{2}{|c|}{of $\sigma_{\Sigma^-}(P^0)$} &
$\sigma_{\Sigma^-}(D^-_s)\simeq 5~ \mu b/N$ \\[1mm]
\multicolumn{2}{|c|}{at $P_{\Sigma^-}=~$650 GeV} & FNAL E781
\\[1mm] \hline \hline
& $\sigma_{\Sigma^-}(P^0)|_{X_f>0}$ & $\sim 50~ -~ 5 ~nb/N$ \\[1mm] \cline{2-3}
$P^0$ & B$_w$ & 0.05 \\[1mm] \cline{2-3}
\quasistable; & $\varepsilon_1$ & $<$~ 0.08  \\[1mm] \cline{2-3}
$X_f>0$ & $\sigma(P^0)_{\rm eff}$ & $<$~0.2~-~0.02~ nb/N \\
[1mm] \cline{2-3}
&$N_w$ &  $<$~ 2800~-~280 events \\[1mm] \hline  \hline
&$\sigma_{\Sigma^-}(P^0)|_{X_f>0.5}$ &  $\sim ~1.5 ~-~0.15~nb/N$  \\[1mm]
\cline{2-3}
$P^0$ &$B_{s} ~  or ~ B_{w}$ &  0.05 \\[1mm] \cline{2-3}
&  $\varepsilon_2$ & $<$~ 0.25 \\[1mm] \cline{2-3}
$X_f>0.5$ & $\sigma(P^0)_{\rm eff}$ &$<$~ $0.02-0.002~nb/N$ \\[1mm] \cline{2-3}
& $N_s ~ or ~ N_w$ &  $<$~ 280-28 events \\[1mm] \hline
\end{tabular} \\
Notes to Table 1:
\begin{enumerate}
\item
Here $\sigma_{\Sigma^-}(P^0)=\sigma(\Sigma^-+N\rightarrow P^0+X)$, etc. $B_{w}$
is the effective branching ratio for visible weak decays of quasistable exotic
Pentaquark baryons:\\ $B_{w}= B[P^0~~\rightarrow~~ p \phi \pi^- + p \phi  \pi^-
\pi^+ \pi^- + p K^{0*} K^-] ~\simeq~0.05$, as described in the text.\\ $B_{s}$
is the effective branching ratio for visible decays of a Pentaquark exotic
baryon resonance: \\ $B_{s}= B[P^0 ~\rightarrow~pD^-_s] \cdot B[D^-_{s}
visible] ~\simeq~(1.0)\cdot(0.05)~\simeq~0.05$.\\ $\sigma(P^0)_{\rm
eff}=\sigma_{\Sigma^-}(P^0)\cdot B\cdot \varepsilon$, where B is either $B_{w}$
or $B_{s}$. We use $\varepsilon_1~<~0.08$ for the average efficiency for the
reconstructed charm events in E781 for $X_f>0$; $\varepsilon_1~<~$(trigger
efficiency)$\cdot$ (reconstruction efficiency)~$<$~(0.18)$\cdot$ (0.45) ~=~0.08
\cite {russ}. The efficiency depends on the lifetime, and also on the
particular decay mode observed, and also on the average X$_f$ value for
Pentaquark events. We use $\varepsilon_2~<~ (0.55)\cdot(0.45)~=~0.25$ for the
same efficiency in the fragmentation region $(X_f>0.5)$; since the trigger
efficiency is higher \cite {russ} for higher $X_f$ particles. We estimate the
Pentaquark production cross section in the fragmentation region (quasi-stable
or resonant) as $\sigma_d \sim 0.03 \cdot \sigma(X_f > 0)$. $N_w$ is the number
of weak decay events for quasistable $P^0$. $N_s$ is the number of decay events
for a resonance $P^0$. The estimated events are extrapolations for a planned
Fermilab E781 1996-97 50 week data run \cite{russ}: $1.3 \times 10^{11}$
interactions in the target, an estimated 2.8$\times 10^8$ charm events, and a
sensitivity of roughly $1.4\times 10^4$~events/(nb/N) of effective charm cross
section. The values cited for N$_w$ and N$_s$ are only projected upper limits,
as described in the text.
\end{enumerate}
\end{table}

\end{document}